\documentclass[multphys,vecphys]{svmult}

\usepackage{makeidx}         
\usepackage{graphicx}        
\usepackage{multicol}        
\usepackage[bottom]{footmisc}

\makeindex             

\begin{document}

\title*{AGN Heating through Cavities and Shocks}
\author{P.E.J. Nulsen\inst{1}, C. Jones\inst{1}, W.R. Forman\inst{1}, L.P.
  David\inst{1}, B.R. McNamara\inst{2,3}, D.A.~Rafferty\inst{3}, L.
  B\^{\i}rzan\inst{3} \and M.W. Wise\inst{4}} 
\authorrunning{Nulsen {\it et al.}}
\institute{Harvard Smithsonian Center for Astrophysics, 60 Garden St
  MS6, Cambridge, MA 02138, USA
\texttt{pnulsen@cfa.harvard.edu}
\and Department of Physics \& Astronomy, University of Waterloo,
  Ontario, Canada
\and Astrophysical Institute and Department of Physics and
  Astronomy, Ohio University, Athens, OH 45701, USA
\and Astronomical Institute ``Anton Pannakoek,'' University of
  Amsterdam, Kruislaan 403, 1098 SJ Amsterdam, The Netherlands}
%
%
\maketitle

\section{Introduction} \label{sec:intro}

\newcommand{\eg}{{\it e.g.}}

The X-ray emitting gas with cooling times much shorter than the Hubble
time in elliptical galaxies, and at the centers of groups and clusters
of galaxies is in an unstable state.  If it is heated at a mean rate
much less than the power it radiates, much of it would cool quickly to
low temperatures, forming reservoirs of cold gas and young stars well
in excess of those that are observed, \eg\ \cite{brm06,rmn06,e01}.  If
the gas is heated at a rate significantly exceeding the power it
radiates, its cooling time would increase until it became comparable
to the Hubble time or longer.  Thus, the relatively high incidence of
short central cooling times requires that the gas is heated at a mean
rate closely matching the power it radiates.  This is difficult to
explain, unless heating rates are coupled to cooling rates.  AGN
heating is a natural vehicle to provide the coupling
\cite{csf02}.

It has been established that the energy released by AGN at the
centers of cooling flows is sufficient to have a significant impact on
the cooling gas \cite{brm04,rmn06}.  Not only are the energies from AGN
outbursts comparable to those needed to stop the gas from cooling, but
the mean powers of the outbursts are well correlated with the powers
radiated.  While the heating process is not well understood, this
would be a remarkable coincidence if AGN heating does not play a
significant role in preventing the gas from cooling.

Three comments on AGN heating of cooling flows are made here.  First,
a simple physical argument is used to show that the enthalpy of a
buoyant radio lobe is converted to heat in its wake.  Thus, a
significant part of ``cavity'' enthalpy is likely to end up as heat.
Second, the properties of the repeated weak shocks in M87 are used to
show that they can plausibly prevent gas close to the AGN from
cooling.  As the most significant heating mechanism at work closest to
the AGN, shock heating probably plays a critical role in the feedback
mechanism.  Finally, results are presented from a survey of AGN heating
rates in nearby giant elliptical galaxies.  With inactive systems
included, the overall AGN heating rate is reasonably well matched to
the total cooling rate for the sample.  Thus, intermittent AGN
outbursts are energetically capable of preventing the hot atmospheres
of these galaxies from cooling and forming stars.

\section{Cavity Heating} \label{sec:cavheat}

\begin{figure}
\centering
\includegraphics[height=3.3truecm]{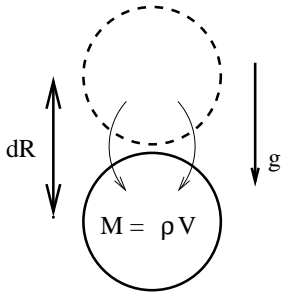}
\caption{Schematic of a buoyantly rising cavity} \label{fig:bubble}
\end{figure}

X-ray decrements over radio lobes are generally consistent with the
lobes being devoid of hot gas, \eg\ \cite{wmn06}, so we treat them as
massless.  From the perspective of the ICM, a cavity (lobe) rises
because ICM falls inward around it to fill the space it occupies
(Fig.~\ref{fig:bubble}).  This converts gravitational potential energy
to kinetic energy in the gas flow around the rising cavity.  Details
of the flow depend on the viscosity, which is poorly determined.  If
it is high, the flow is laminar and the kinetic energy is dissipated
as heat over a region comparable in size to the cavity (akin to
Stoke's flow around a sphere \cite{ll59}).  If the viscosity is very
low, the Reynolds number would be high and the flow turbulent.  The
turbulent region near the cavity would have a similar size to it.
Turbulent kinetic energy is dissipated in the turn over time of the
largest eddies \cite{ll59}, so that the dissipation time $t_{\rm d}
\sim r_{\rm cav} / v_{\rm cav}$, where $r_{\rm cav}$ is the radius of
the cavity and $v_{\rm cav}$ is the speed of the eddy, which is
comparable to the speed of the cavity.  Since $t_{\rm d} v_{\rm cav}
\simeq r_{\rm cav}$, much of the turbulent kinetic energy is
dissipated in a region of comparable size to the cavity.  Thus,
regardless of the viscosity, the kinetic energy created by cavity
motion is dissipated locally as heat in a wake of similar size to the
cavity.

\newcommand{\etal}{{\it et al.}}
\newcommand{\ie}{{\it i.e.}}

The gravitational potential energy that is released as the cavity
rises a small distance $dR$, subject to the gravitational acceleration
$g$ is (Fig.~\ref{fig:bubble})
\begin{equation} \label{eqn:cavheat}
M g \, dR = V \rho g \, dR \simeq - V {dp \over dR} dR = - V \, dp,
\end{equation}
where $M = \rho V$ is the mass of gas displaced by the cavity, $V$ is
the volume of the cavity and $\rho$ is the density of the external
gas.  Cavity motion is generally subsonic, so that the gas around the
cavity is approximately hydrostatic, \ie, $\rho g \simeq - dp/dR$.
The final $dp$ here is the pressure change in the external gas over
the distance $dR$, but a subsonic cavity maintains approximate local
pressure equilibrium, so that $dp$ may also be regarded as the change
in pressure of the cavity.  In terms of the enthalpy, $H = E + p V$,
the first law of thermodynamics, $dE = T\, dS - p \, dV$, becomes $dH
= T\, dS + V \, dp$.  In an adiabatic process, the heat exchange, $T
\, dS$, is zero and $dH = V \, dp$.  Thus, equation
(\ref{eqn:cavheat}) states that the potential energy released as the
cavity rises is equal to the decrease in its enthalpy.  No exotic
process is required to explain how cavity enthalpy is converted to
heat.  This argument is less accurate for the largest cavities (with
$r_{\rm cav} \simeq R$), but the corrections are of order unity.
Fragmentation of cavities does not change the result, unless 
cavity contents dissipate.  We conclude that enthalpy lost by rising
cavities is dissipated as heat locally in their wakes \cite{brm04}.

\section{Weak Shock Heating}

Weak shocks associated with AGN outbursts are generally only detected
in deep X-ray observations, so that relatively few are known, \eg\
\cite{fnh05,mnw05,nmw05,fst06,mnj06}, although they are probably
produced in most AGN outbursts.  The energies required to drive the
shocks are comparable to cavity enthalpies, but much of this energy
will end up as potential energy in the ICM.  The main requirement for
stopping gas from cooling is to replace the entropy lost by radiation
and, since the entropy jump, $\Delta S$, varies as the cube of the
pressure jump, weak shocks are not very effective at this
\cite{dnm01}.  The equivalent heat input per unit mass can be
evaluated as  
\begin{equation} \label{eqn:sjum}
\Delta Q \simeq T \Delta S = E \Delta \ln {p \over \rho^\gamma}, 
\end{equation}
where $E$ is the specific thermal energy, $p$ the pressure and $\rho$
the density of the gas.  Thus, the fractional heat input, $\Delta Q /
E$, is given by the jump of $\ln p/\rho^\gamma$ in the shock.

Three weak shocks are visible in the X-ray image of M87 \index{M87}
\cite{fcj06}.  Fitting a simple model to the surface brightness
profile of the innermost shock at 0.8 arcmin ($\simeq 3.7$ kpc) gives
a Mach number of $\simeq 1.4$ and a shock age of $\simeq
2.4\times10^6$ y.  Its equivalent heat input (equation
\ref{eqn:sjum}), determined from the Mach number, is $\Delta Q / E
\simeq 0.022$.  Clearly, this single shock does very little to heat
the gas.  However, there is a second shock at about twice the radius,
and the third shock, at $\simeq 3$ arcmin, required several times more
energy \cite{fnh05}.  Thus, a shock of comparable strength to the 0.8
arcmin shock may well occur every $\sim 2.5\times10^6$ y.  The cooling
time of the gas at 0.8 arcmin $\simeq 2.5\times10^8$ y, so that there
is time for $\sim 100$ such shocks during the cooling time.
Therefore, the combined heat input of the shocks ($100 \times 0.022 =
2.2$) is more than enough to make up for radiative losses from the
gas.

These numbers are not to be taken literally, but they make a
plausible case that the modest heating of repeated weak shocks can
stop the gas at 0.8 arcmin from the AGN in M87 from cooling.  The
cavity heating discussed in section \ref{sec:cavheat} is only
effective beyond the radius 
where the radio lobes form.  Some other mechanism is required to
prevent gas closer to the AGN from cooling.  Since the gas closest to
the AGN is likely to be its most significant source of fuel, the
heating process that affects this gas probably plays a critical role
in the feedback cycle that prevents gas from cooling.  Thus, while
weak shock heating is unlikely to be the dominant mode \cite{frt05},
it may well play a major role in the AGN feedback process.

\section{AGN Heating in Nearby Elliptical Galaxies}

\begin{figure}
\centering
\includegraphics[height=6truecm,angle=270]{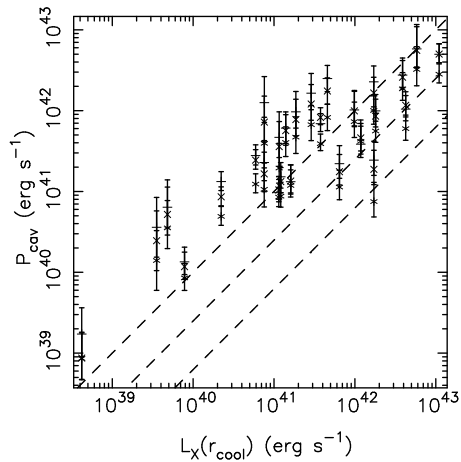}
\caption{Cavity heating power vs cooling power for nearby elliptical
  galaxies.  Heating powers, $P_{\rm cav}$, are estimated as $p V /
  t_{\rm a}$, using three different estimates of the age, $t_{\rm a}$,
  for the 27 galaxies with cavities in the sample of Jones \etal.
  Cooling power is the X-ray luminosity from within the projected
  radius where the cooling time equals $7.7\times10^9$ y.  The dashed
  lines show where heating power equals cooling power, for heat inputs
  per cavity of $pV$, $4pV$ and $16pV$, from top to bottom.}
  \label{fig:ecavpow}
\end{figure}

To date, deeper X-ray observations of clusters have almost invariably
revealed more cavities, \eg\ \cite{fcj06}, so that heating powers are
likely to be underestimated from existing data \cite{brm04,rmn06}.
Progress in this areas is also hampered by relatively poor
understanding of the selection effects for finding cavities
\cite{brm04}.  A deep survey of a complete sample of cooling flow
clusters is required to assess the overall significance of AGN heating
in clusters.  In the mean time, Jones \etal\ (in preparation and this
proceedings) have assembled X-ray observations of a nearly complete
sample of nearby giant elliptical galaxies.  This includes
$\simeq 160$ galaxies, 109 of which show significant diffuse emission
from hot gas after removal of resolved and unresolved point sources.
Of those, 27 have significant AGN cavities.  The sample is used here
to assess the overall significance of AGN heating.

Our determinations of AGN heating rate parallel those of B\^{\i}rzan
\etal\ \cite{brm04} and Rafferty \etal\ \cite{rmn06}, with minor
modifications.  Where gas temperatures are not available from the
literature, they are estimated from the velocity dispersion, $\sigma$,
of a galaxy as $kT = 1.5 \mu m_{\rm H} \sigma^2$ (consistent with the
median of $\mu m_{\rm H} \sigma^2 / (kT)$ for the remaining galaxies).
For the three galaxies without a temperature or velocity dispersion,
the gas temperature is set to the median value of 0.7 keV.  Abundances
are assumed to be 0.5 solar.  Electron densities are determined from
beta model fits.  Heating powers are estimated as $pV$ per cavity,
divided by one of the three estimates of cavity age, the sound
crossing time, the buoyant rise time, or the refill time \cite{brm04}.
The cooling power is taken as the X-ray luminosity from within the
projected radius where the cooling time equals $7.7\times10^9$ y.
Our cooling powers are not corrected for mass deposition, {\it cf.}
\cite{brm04,rmn06}.

The three resulting estimates of heating power are plotted against the
cooling power in Fig.~\ref{fig:ecavpow}.  Note that for UGC~408,
\index{UGC 408} the AGN appears to lie near the center of a single
cavity, suggesting that the system is viewed almost along its radio
axis, with its cavities and AGN projected on top of one another.  The
apparent distance from the AGN to the center of the cavity is then
much smaller than the real distance, causing the age of the outburst
to be underestimated and the heating power to be overestimated.
Consistent with this, the naive estimates of its heating power are
exceptionally high.  For UGC 408 alone, we have therefore assumed
that the distance from the center of the cavity to the AGN is equal to
the semimajor axis of its cavity.  The corrected heating powers agree
with those for other systems with similar cooling powers.  This
correction reduces our estimate of the total heating power for all
systems by a factor of $\sim2$.

Approximately half of the outbursts in cooling flow clusters have
heating powers that match or exceed their cooling powers for a heat
input of $4pV$ per cavity \cite{rmn06}.  By contrast, this is true for
all of the nearby giant ellipticals in our sample
(Fig.~\ref{fig:ecavpow}), apart from NGC 1553 \index{NGC 1553}
\cite{bsi01}.  Conversely, only about one quarter of the giant
ellipticals with significant emission from hot gas have cavities,
whereas the fraction of cluster cooling flows with outbursts is closer
to 70\% \cite{dft05}.  Consistent with the AGN feedback model, this
suggests that the duty cycles of outbursts in larger systems have to
be greater to keep the mean heating power at the level needed to stop
cooling.

Allowing $1 pV$ per cavity, the total heating powers for the 27
systems are $2.6\times10^{43}$, $2.9\times10^{43}$ and
$1.5\times10^{43}\rm\ erg\ s^{-1}$, corresponding to ages of the sound
crossing times, the buoyant rise times and the refill times,
respectively.  The total cooling power for all 109 galaxies with
significant emission from diffuse hot gas is $1.06\times10^{44}\rm\
erg\ s^{-1}$, so that the ratios of total cooling power to total
heating power for the three age estimates are 4.1, 3.6 and 6.9, in
turn.  The enthalpy of a cavity dominated by relativistic gas is $4pV$
and this value might, typically, be doubled by the ``shock energy''
(more precisely, the $p\, dV$ done as the cavity was inflated).  While
there is still significant systematic uncertainty in these numbers,
they make a good case that AGN outbursts in nearby giant elliptical
galaxies can prevent their X-ray emitting gas from cooling and forming
stars.  The outbursts are intermittent, with the AGN active $\sim
25\%$ of the time.

\smallskip

This work was partly supported by NASA grant NAS8-01130.



\printindex
\end{document}